\documentclass{article}
\usepackage{graphics,epsfig,amsfonts,amsmath}
\begin{document}
\newtheorem{theorem}{Theorem}[section]
\newtheorem{lemma}[theorem]{Lemma}
\newtheorem{proposition}[theorem]{Proposition}
\newtheorem{corollary}[theorem]{Corollary}
\newenvironment{proof}[1][Proof]{\begin{trivlist}
\item[\hskip \labelsep {\bfseries #1}]}{\end{trivlist}}
\newenvironment{definition}[1][Definition]{\begin{trivlist}
\item[\hskip \labelsep {\bfseries #1}]}{\end{trivlist}}
\newenvironment{example}[1][Example]{\begin{trivlist}
\item[\hskip \labelsep {\bfseries #1}]}{\end{trivlist}}
\newenvironment{remark}[1][Remark]{\begin{trivlist}
\item[\hskip \labelsep {\bfseries #1}]}{\end{trivlist}}
\newcommand{\qed}{\nobreak \ifvmode \relax \else
      \ifdim\lastskip<1.5em \hskip-\lastskip
      \hskip1.5em plus0em minus0.5em \fi \nobreak
      \vrule height0.75em width0.5em depth0.25em\fi}
\newcommand{\N}{\mathbb{N}}
\newcommand{\R}{\mathbb{R}}
\newcommand{\Q}{\mathbb{Q}}
\title{Fractal growth of tumors and other cellular populations: linking the mechanistic to the phenomenological modeling and vice versa\footnote{I wish to dedicate
this work to my friends and coworkers Prof. Alessandro Bertuzzi and
Prof. Alberto Gandolfi, who are among the best people I know, and
not only in the scientific world. Of course, they also helped me in
revising the draft of this manuscript, and I thank them very
much.}.}
\author{Alberto d'Onofrio $^{1}$
} 
\date{ July 7 2006}
\maketitle
%
{\center $^1$ Division of Epidemiology and Biostatistics, European Institute of Oncology, Via Ripamonti 435, Milano, Italy, I-20141\\E-mail: donofrio@mail.dm.unipi.it, Phone:+390257489819\\[3mm]  }
\begin{abstract}
In this paper we study and extend the mechanistic mean field theory
of growth of cellular populations proposed by Mombach et al in
[Mombach J. C. M. et al., \textit{Europhysics Letter}, \textbf{59}
(2002) 923] (\emph{MLBI} model), and we demonstrate that the
original model and our generalizations lead to inferences of
biological interest. In the first part of this paper, we show that
the model in study is widely general since it admits, as particular
cases, the main phenomenological models of cellular growth. In the
second part of this work, we generalize the \emph{MLBI} model to a
wider family of models by allowing the cells to have a generic
unspecified biologically plausible interaction. Then, we derive a
relationship between this generic microscopic interaction function
and the growth rate of the corresponding macroscopic model. Finally,
we propose to use this relationship in order to help the
investigation of the biological plausibility of phenomenological
models of cancer growth.

\end{abstract}

\section{Introduction}
Human beings, animal and plants are sets of cells. As a consequence,
modeling the growth of cellular populations should be considered
among the most important scientific topics. Surprisingly, the vast
majority of theoretical work on this subject \cite{Ara} produced
essentially mathematical models based on qualitative "macroscopic"
biological reasoning or, as in the case of the well known Gompertz
model, purely on the ground of good fit to experimental data
\cite{Ara}. Among the few works aimed at introducing a mechanistic
theory in order to link macroscopic to microscopic parameters, the
model (hereafter referred as the \emph{MLBI} model) proposed by
Mombach et al. \cite{M} is particularly interesting because of its
simplicity and biological plausibility, being based on the realistic
hypothesis of long range interactions between cells in a population
whose "structure is a fractal" \cite{M}. Furthermore, relying on the
fractal structure of cells aggregates, is adequate to describe tumor
growth \cite{N,WK1}, at least in some particularly relevant
biological frameworks. As a consequence of its biological realism,
\emph{MLBI} model allowed its authors to unify at microscopic level
three well known growth laws (logistic, gompertzian and exponential
laws \cite{Bajzer}). In other words, apparently contradictory growth
models  are simply macroscopic different manifestations of a common
physical microscopic framework (note that at macroscopic level it is
easy to show that those models are linked). However, many other
relevant growth laws were proposed. Among them we cite the von
Bertanlaffy's \cite{Bajzer}, the West-Guiot law \cite{Gu1,Gu2}, the
power law \cite{Agur,Drasdo,Bru}, all of which well describe
experimental "in vitro" and "in vivo" data \cite{Bajzer, Agur, Bru}.

This "proliferation" of different models, all, at least apparently,
theoretically sound, and with some encouraging experimental
evidences, should not surprise the non biologist reader. For
example, let us consider solid tumors (to which we shall mainly
refer in the following): with the term "solid tumors" one summarizes
a wide range of polymorphic diseases characterized by at least three
levels of growth behaviors: 1)disease-specific level: different
kinds of tumors may exhibit different kinds of growth; 2)the
inter-patients level: two similar subjects, suffering of the same
kind of cancer, may show very different time courses of the
diseases; 3)intra-patient level: the main tumor may have growth
characteristics that are different from those of its metastases.
Finally, there is also a temporal level: tumor is a dynamic disease
whose characteristics of proliferation may change reflecting changes
in its topological structure \cite{Gu2}.

The starting point of this research has been the following question:
given the complexity of growth phenomena of tumor and non-tumor cell
populations, are the encouraging results obtained by many and
different mathematical models simply due to their fitting
flexibility or there is some unifying mechanism ?

Aim of this letter is to show that the \emph{MLBI} model is able to
unify also the above listed important models of growth, and others,
besides those considered in \cite{M}. Thus, we shall show that
apparently very different behaviors in very similar conditions
simply reflect slight differences in physically meaningful (and, in
principle, experimentally measurable) parameters such as the fractal
dimension of the tumor. Analyzing the relationships between the
\emph{MLBI} model and the model \cite{Gu2} we shall also shortly
examine the case of time varying topological properties.

The relevance of our findings, in our opinion is twofold. In fact,
on one hand, the fact that the \emph{MLBI} model may particularize
itself to all main empirical models, confirmed by good fitting of
experimental data, gives more biological soundness to the
\emph{MLBI} model; on the other hand, those models have now a
theoretical ground which might explain them at microscopic level.

Furthermore, we shall extend the \emph{MLBI} model by allowing very
general laws of cell-cell interaction and, in so doing, transforming
it in a meta-model, that unifies many other models, and, starting
from the micro level, to define new ones. We propose a simple method
of "reverse engineering", which might allow insights on the
biological basis of a given phenomenological model of growth of
cellular populations. In fact, we shall show that our generalization
of the \emph{MLBI} model seems to indicate that, for a
macroscopically defined model, its population level growth rate
(\emph{PLGR}) must be not only decreasing with the population size,
as it is well known, but it must also be convex (its second
derivative must be positive).
\section{Phenomenological models and the \emph{MLBI} model}
A phenomenological model that describes the growth of a population
of cells may be written as:
\begin{equation}
    x'= x R(x) \label{Family}
\end{equation}
where $x$ is the size of the population. $R(x)$ denotes the
\emph{PLGR}, which, due to evident biological reasons, must be a
decreasing function of $x$.

Coming to the \emph{MLBI} theory, the assumptions on which the model
is built (other than the above mentioned long range cell-cell
interactions and the fractal spatial structure of the population)
are the following: i)a sufficiently high supply of nutrients (as a
consequence the \emph{MLBI} model is particularly adequate to
describe the behavior of an "in vitro" sufficiently nourished
multicellular spheroid and also "in vivo" tumors in the angiogenic
phase); ii)the inhibiting chemicals are such that they diffuse in
the cellular structure; iii)the individual "cell level"
proliferation rate of a cell results from its baseline replication
rate $G_i$ minus the long range inhibition interactions with all
other cells of the population. For the generic "i-th" cell it is
\cite{M}:
\begin{equation}
R_i = G_i -
\sum_{j=1}^{n}{J\frac{1-\delta_{i,j}}{|\vec{r_i}-\vec{r_j}|^\gamma}}
\label{Fund2}
\end{equation}
where: $J$ is a constant related to the effectiveness of the
inhibitory action, $|\vec{r_i}-\vec{r_j}|$ is the distance between
cell $i$ and cell $j$, $\delta_{i,j}$ is the Kroeneker's symbol. The
function $J |\vec{r_i}-\vec{r_j}|^{-\gamma}$ is an ansatz function,
physically motivated in \cite{M} by the theory of diffusion in
fractal structures \cite{C}. Starting from ~(\ref{Fund2}) and from
the assumption of fractal structure (with fractal dimension $D_f$),
in \cite{M} the following relation has been obtained:
\begin{equation}
   R(x)= \overline{G} - J I(x)  \label{FamilyMecha}
\end{equation}
where $\overline{G}$ is the average of the $G_i$s and:
\begin{equation}      \label{I}
   I(x) =
   \frac{\omega}{D_f -\gamma}\Big( \big(\frac{D_f}{\omega}x\big)^{1-\gamma/D_f}-1\Big) \textrm{
   if } \gamma \neq D_f \textrm{,  }  I(x) =
   \frac{\omega}{D_f}Log\big(\frac{D_f}{\omega}x\big)\textrm{
   if } \gamma = D_f,
\end{equation}
where the constant $\omega$, defined in \cite{M}, is related to the
density of the cellular aggregate. Thus, starting from microscopic
cell-cell interactions, a macroscopic but "non phenomenological"
equation $x' = x R(x)$ has been obtained. Note that, as previously
stressed in \cite{M}, the parameters $J$ and $\gamma$ might be
determined by concentration measurements, and $D_f$ by the pattern
of the cellular structure.

The family of ODE models $x'=x R(x)$ with \emph{PLGR} given by
~(\ref{FamilyMecha}) and ~(\ref{I}) has been studied in \cite{M}
with the following results: for $\gamma = D_f$ the model reduces to
the Gompertz model; for  $0 \ge \gamma < D_f $ the \emph{MLBI} model
gives the generalized logistic model $x' = k_1 x - k_2 x^{\nu}$ with
$\nu = 2 - (\gamma/D_f) \le 2$. However, $\nu =2 \Rightarrow
\gamma=0$ i.e. biologically unrealistic constant interactions
\cite{M}. We add here that this non-realism might explain the poor
performances of the logistic model in fitting experimental tumor
data, as reported in \cite{Bajzer}.
\section{The general behavior for $\gamma > D_f$}
Unfortunately, because of a minor error in \cite{M}, the exploration
of the parameter space was not deepened as it deserved. In fact, the
authors claimed that for $\gamma
> D_f$ the behavior of the system is of exponential type. We
detected the trivial error and discovered that all the other main
phenomenological models of growth may be considered as particular
cases of the \emph{MLBI} model.

If $\gamma > D_f$, let us set $\gamma = D_f(1+Q)$ with $Q>0$, that
yields:
\begin{equation} \label{rgq}
    R(x)= \frac{J}{Q}\big(\frac{\omega}{D_f}\big)^{Q+1}x^{-Q} + \overline{G}- \frac{J \omega}{D_f
    Q}.
\end{equation}
Note that the above \emph{PLGR} is decreasing and also convex:
\begin{equation}
    R'(x)= - J \big(\frac{\omega}{D_f}\big)^{Q+1}x^{-Q-1} < 0 \textrm{, }
    R''(x)= (Q+1) J \big(\frac{\omega}{D_f}\big)^{Q+1}x^{-Q-2} > 0  \label{rgq3}
\end{equation}
We shall show in the final section that the convexity is as
important as the negativity of the derivative $R'(x)$.

The solution of $x'=x R(x)$ when :
\begin{equation}
   \overline{G}- \frac{J \omega}{D_f Q} <0 \Rightarrow    D_f < \gamma < D_f + \frac{\omega J}{\overline{G}} \label{diseq}
\end{equation}
is not the exponentially increasing one given in \cite{M}, since the
equation $x'=x R(x)$ has a global attractor:
\begin{equation}
  lim_{t \rightarrow +\infty } x(t;x(0)) = x_{eq} = \frac{\omega}{D_f}\Big(\frac{J \omega}{J \omega - \overline{G} Q D_f}\Big)^{1/Q} = \frac{\omega}{D_f} \Big(\frac{J \omega}{J \omega - \overline{G}\gamma + \overline{G} D_f}\Big)^{D_f/(\gamma-D_f)} \label{ciccio}
\end{equation}
and the dynamics of $x(t)$ is as follows:
\begin{equation}
  x(t) =\Big( x_{eq}^Q  +\big(x(0)^Q-x_{eq}^Q\big)Exp\big( -t*(-\overline{G}Q + \frac{J \omega}{D_f }) \big) \Big)^{1/Q} \label{xciccio}
\end{equation}
On the contrary, if $\gamma > \gamma^* := D_f + \omega
J/\overline{G} $ there is exponential explosion of the growth:
\begin{equation}
  x(t) \propto Exp\Big(\big(\overline{G}- \frac{J \omega}{\gamma-D_f}\big)t\Big)  \label{espo}
\end{equation}
In view of some experimental findings, exponential growth has been
traditionally associated to initial stages of growth. Thus
eq.(\ref{espo}) may be read as follows: the case $\gamma > \gamma^*$
is not likely to be observed or it is related to very quickly
growing tumors for which a long temporal observation is, of course,
not possible.
\section{Power law growth} Recently,  by means of a
phenomenological approach and without explicitly assuming the
presence of a necrotic core, Hart et al \cite{Agur} formulated an
interesting model whose solution is a power law growth of the tumor
mass, and validated this model by means of fitting to breast cancer
data. A similar behavior was also predicted in \cite{Drasdo}, where
an interesting individual based model was used, and in \cite{Bru},
where also experimental data are reported (see also \cite{Ara}). So
an important feature for a unified mechanistic model would be to
allow such a behavior. Actually, this happens in our case, since if:
\begin{equation}
   \overline{G}- \frac{J \omega}{D_f Q} = 0 \Rightarrow \gamma = D_f +  \frac{J \omega}{\overline{G}} \label{eq0}
\end{equation}
the tumor will grow following an asymptotically power law. In fact,
when ~(\ref{eq0}) holds:
\begin{equation}
x'= \frac{J}{Q}\big(\frac{\omega}{D_f}\big)^{Q+1}x^{1-Q} \Rightarrow
x(t) = \Big( x_o^{J\omega/(D_f \overline{G})} + J
\big(\frac{\omega}{D_f}\big)^{1+J\omega/(D_f \overline{G})} t
\Big)^{D_f \overline{G}/(J \omega )}. \label{soleqd0}
\end{equation}
For large times:
\begin{equation}
x(t) \propto t^{\frac{\overline{G}}{J \omega}D_f }
\label{soleqd0largetimes}
\end{equation}
Note that the power depends on the fractal features of the aggregate
of growing cells. In particular, linear growth of the diameter of
the spheroid, corresponding to cubic variation of the cell number
(quadratic in 2D), is obtained when $\gamma = 4 D_f/3$ (in 2D:
$\gamma = 3 D_f/2$). According to \cite{Agur}, the power law growth
we have found does not depend on the localization of the
proliferating cells near the surface of the spheroid. We remark that
our analysis predicts the possibility of power law growth only for a
specific combination of the parameters. This might be read in a
dichotomic way: either the model has to be modified to manage the
power law more robustly, or the power law may be read as a limit
case. The same considerations hold for the Gompertz model.

Summarizing the findings of this and of the previous sections, we
may state that $\gamma$ is a bifurcation parameter with threshold
value $ \gamma^*= D_f + J \omega/\overline{G}
> D_f$ determining a "catastrophic" transition, since $x(t)$ is bounded for $\gamma
< \gamma^*$, whereas the growth is unbounded for $\gamma \ge
\gamma^*$. The threshold for the unbounded growth depends not only
on the geometrical parameter $D_f$, as stated in \cite{M} (where the
threshold was $D_f$), but also on the parameter $J$ (with $\partial
\gamma^* / \partial J > 0$) and on the proliferation parameter
$\overline{G}$ (with $\partial \gamma^* /
\partial \overline{G}  < 0$). Thus, higher values of the
proliferation parameter imply a  lower threshold for the spatial
decaying $\gamma$, whereas high values of $J$ require higher values
for $\gamma$ to have power law or exponential expansion.
\section{Linear growth law and linear growth} For $Q=1$, i.e.
 $\gamma = 2 D_f$, the law of the growth $x'=x
R(x)$ becomes  a linear first order equation:
\begin{equation}
x' = J \big(\frac{\omega}{D_f}\big)^{2} + \big(\overline{G}- \frac{J
\omega}{D_f}\big)x. \label{linode}
\end{equation}
To the best of our knowledge, this is the first time that the
possibility of such a simple law of growth is contemplated in the
context of tumor growth and with a  plausible mild constraint on
parameters. If $\overline{G}- J \omega / D_f < 0$ the tumor size
will be attracted to the equilibrium point:
\begin{equation}
x_{eq} = \frac{J \omega^2}{D_f(J\omega - \overline{G} D_f)}
\label{epl}
\end{equation}
or, if $\overline{G}- J \omega / D_f >0$, it will expand
exponentially.  If $\overline{G}- J \omega / D_f = 0$, the growth of
the tumor is linear from $t=0$:
\begin{equation}
x(t) = x(0) + J \big(\frac{\omega}{D_f}\big)^{2} t. \label{lin}
\end{equation}
\section{Comparison with the model by del Santo, Guiot et al. \cite{Gu1,Gu2}} We show here that the very interesting model
of growth proposed by del Santo, Guiot and coworkers in
\cite{Gu1,Gu2} is a particular case of the \emph{MLBI} model. In
fact the model in \cite{Gu1} reads as follows:
\begin{equation}
x' = x \big(a x^{p-1} - b\big)  \label{guiot}
\end{equation}
with $p \in (2/3,1)$, $a>0$  and $b>0$. Matching with ~(\ref{rgq});
\begin{equation}
p = 2-\frac{\gamma}{D_f}  \label{pdf}
\end{equation}
which leads to :
\begin{equation}
 1 < \frac{\gamma}{D_f} < \frac{4}{3}\textrm{, } \frac{\omega J}{\overline{G}} >  \frac{D_f}{3}   \label{dfgammalimits}
\end{equation}
For tridimensional tumors following the Guiot's law this implies
that it must be $\gamma <4$.

Up to now we assumed that the topological properties of the tumor
were constant. On the contrary, as Guiot et al stressed in their
paper, these properties, and namely the fractal dimension, change
during tumor development, in a way that the scaling exponent $p$ is
increasing at least for well perfused tumors in the angiogenic phase
(i.e. when there is no lack of nutrients, as required for the
rigorous application of the \emph{MLBI} model). This means to assume
an increasing $D_f(t)$ or, if we assume that also $\gamma$ is a time
function, to assume that $D_f(t)$ increases faster than $\gamma(t)$.
Finally, tumor for which data fitting shows constant $p$ should have
$\gamma(t) \propto D_f(t)$.
\section{The von Bertanlaffy's model} One among the early
mathematical models of growth, due to von Bertanlaffy \cite{Bajzer},
was as follows:
\begin{equation}
 x' = x (a x^{-1/3} - b) \label{vonB}
\end{equation}
and it is easily recovered as a particular case of the \emph{MLBI}
model provided that:
\begin{equation} \gamma = \frac{4}{3} D_f \textrm{, } \frac{\omega
J}{\overline{G}} > \frac{D_f}{3}  \end{equation}
\section{Relationships between $\gamma$ and $D_f$}
We saw that for $\gamma = D_f + (J \omega)/\overline{G} $ there is
power law growth, and for higher values of the fractal dimension the
growth is more rapid, which well agrees with usual tumor behavior
where a high $D_f$ is considered an index of high growth velocity
(and, as a consequence, of poor survival) \cite{N,D}. However, for
$D_f < \gamma < D_f + (J \omega)/\overline{G} $, if we use eq.
(\ref{ciccio}) we obtain that: $\partial_{D_f} x_{eq}(D_f,\gamma)
<0$, and when $\gamma > D_f + (J \omega)/\overline{G}$ it is
$\partial_{D_f} x(t;D_f,\gamma) <0$ . Apparently, the model seems to
fail in reproducing an important biological property. However, we
note that in the Guiot's and in the von Bertanlaffy's case, where
there is a linear relationship between $\gamma$ and $D_f$, it is
$(d/d D_f) x_{eq}(D_f,\gamma(D_f)) > 0$. Similarly, for $\gamma
> \gamma^*$, one may find linear relationships such that the
exponential grows faster. Finally, for for $\gamma<D_f$ for whatever
$\gamma=h(D_f)$ with $h'(D_f)>0$ it is: $(d/d D_f)
x_{eq}(D_f,h(D_f)) < 0$. Summarizing these findings, it seems that
it should be $\gamma > D_f$ and that $\gamma$ and $D_f$ must be
positively correlated.
\section{Extending the \emph{MLBI} model}
We may observe that the inhibition function
$|\vec{r_i}-\vec{r_j}|^{-\gamma}$ used in ~(\ref{Fund2}), may assume
very high values in the neighboring of the $i$ cell.  Furthermore,
that particular ansatz was chosen in \cite{M} because of the
assumption of diffusion of inhibiting chemicals, which might be no
more valid in case of well vascularized tumors. Thus, we change the
function $J |\vec{r_i}-\vec{r_j}|^{-\gamma}$ with another generic
positive and decreasing decreasing ansatz function
$F(|\vec{r_i}-\vec{r_j}|)$ ($F(u)>0$, $F'(u)<0$, $F(+\infty)=0$ and,
if it is necessary, $F(0) < +\infty$). One has:
\begin{equation}
R_i = G_i - \sum_{j=1}^{n}(1-\delta_{i,j})F(|\vec{r_i}-\vec{r_j}|)
\label{Fund3}
\end{equation}
Since the spheroid radius is related to $x$ by the following formula
(ref. \cite{M}, pag. 925): $R_{max} = (x D_f /\omega)^{1/D_f}$,  and
proceeding as in \cite{M} but using $F(r)$ instead of the specific
ansatz $J r^{-\gamma}$, we easily obtain a generalized \emph{PLGR},
$R_{gen}$:
\begin{equation}
R_{gen}(x) =   \overline{G} - \omega
\int_{ro}^{R_{max}(x)}{F(r)r^{D_f-1}dr},  \label{Rgen}
\end{equation}
which defines a very general family of models of growth. This family
is such that all the models belonging to it share two important
properties: $R_{gen}(x)$ is decreasing and convex:
\begin{equation}
R_{gen}'(x) =  - F\Big(\big(\frac{D_f}{\omega} x \big)^{1/D_f}\Big)
< 0 \textrm{  ; } R_{gen}''(x) =  -\frac{1}{D_f}
\Big(\frac{D_f}{\omega} \Big)^{1/D_f} x^{-1+1/D_f}
F'\Big(\big(\frac{D_f}{\omega} x \big)^{1/D_f}\Big)
> 0 \label{d2Rgen}
\end{equation}
Note that:
\begin{equation}
Signum\big(R_{gen}''(x) \big)  \Leftrightarrow   -
Signum\big(F'(x^{1/D_f})\big)  \label{iff}
\end{equation}
We think that the relationship ~(\ref{iff}) has implications of
relevant biological interest. In fact, as long as the basic
hypotheses stated in \cite{M} are valid, the $\Leftarrow$
relationship means that the assumption of bio-physically realistic
interaction between the cells (i.e. interactions decreasing with the
distance) implies that the \emph{PLGR} of the corresponding mean
field macroscopic growth model is convex. On the contrary, a concave
\emph{PLGR} would correspond to a unrealistic $F$. The $\Rightarrow$
relationship may be read as follows: given a macroscopic growth
model, if its \emph{PLGR} is convex, then the underlying microscopic
cell-cell interaction is decreasing, which is biologically
meaningful. On the contrary, if $R''(x)<0$ then it would correspond
to an unrealistic cell-cell interaction increasing as the distance
between the cells increases.

More explicitly, we propose the study of the convexity of $R(x)$
(i.e. the sign of $R''(x)$)as a way to \textit{start} investigating
the biological plausibility of a phenomenological cellular growth
model, and to classify a model characterized by negative or varying
sign $R''(x)$  at least as "suspect". An example is the well known
generalized logistic model with exponent $\nu > 2$: $ x' = a x - b
x^{\nu}$. This model has been introduced in \cite{Ayala}, in a
context (populations of insects) where values $\nu \ge 2$ are
perfectly biologically plausible. Analyzing its \emph{PLGR} $R(x)=a
- b x^{\nu-1}$, it is easy to see that this kind of logistic model
would correspond to a non-realistically increasing cell interaction
function: $ \nu
>2  \Rightarrow  \gamma <0 \Rightarrow F'(r)=(-\gamma)r^{-\gamma-1}
>0$. A different example is the following: let us consider an
exponentially decreasing \emph{PLGR}: $ R(x) = A Exp\big(-B x^d\big)
- C \textrm{, } 0<d \le 1$. This \emph{PLGR} is decreasing  ($R'(x)
<0$) and convex ($R''(x)>0$), as it is easy to verify. Furthermore,
it would correspond to a decreasing interaction function $F(r) =
r^{-D_f(1-d)}Exp\big(- q r^{d*D_f}\big)$, which may be read as the
$r^{-\gamma}$ function damped by a decreasing exponential. For $d=1$
we have the particular case $F(r)=Exp\big(- q r^{D_f}\big)$.
Finally, let us consider the Wheldon model \cite{Wjtb} in which
$R(x)=-w +a/(1+b x)$, which, not surprisingly, is found to be convex
($R''(x)>0$) and which, as it is easy to verify, is related to the
inhibition function $F(r)=1/(1+ \hat{b} r^{D_f})^2$.
\section{Concluding remarks} Our extension of \cite{M}, aimed to
stress the role of the cell-cell interactions, neglects important
phenomena and may be further extended to model them.  For example,
the interaction with the microenvironment may be encoded in the
parameters $\overline{G}$ and $J$, either as noisy fluctuations or
as deterministic dependence on new state variables (e.g. the density
of blood microvessels). Some extensions might be less trivial, e.g.
introducing subpopulations of quiescent and necrotic cells. This
extension would be important in order to relax the basic hypothesis
\cite{M} of a sufficiently high level of nutrients, and to include
the Gompertz law and the power law growth in a more robust way.

\end{document}